\documentclass[twocolumn,english,aps,prl,floats,showpacs,amsmath,amssymb,floatfix]{revtex4}
\usepackage[T1]{fontenc}
\usepackage[latin9]{inputenc}
\usepackage{amsmath}
\usepackage{amssymb}
\usepackage{graphicx}
\usepackage{esint}

\makeatletter
\@ifundefined{textcolor}{}
{%
 \definecolor{BLACK}{gray}{0}
 \definecolor{WHITE}{gray}{1}
 \definecolor{RED}{rgb}{1,0,0}
 \definecolor{GREEN}{rgb}{0,1,0}
 \definecolor{BLUE}{rgb}{0,0,1}
 \definecolor{CYAN}{cmyk}{1,0,0,0}
 \definecolor{MAGENTA}{cmyk}{0,1,0,0}
 \definecolor{YELLOW}{cmyk}{0,0,1,0}
}

\@ifundefined{definecolor}
 {\usepackage{color}}{}
\usepackage{epsfig}\usepackage{dcolumn}
\usepackage{bm}

\makeatother

\usepackage{babel}
\begin{document}

\title{Dynamic response of a single-electron transistor in the ac Kondo
regime}

\date{\today}

\author{Thanh Thi Kim Nguyen}

\address{Department of Physics, University of Cincinnati, Cincinnati, Ohio
45221, USA}
\begin{abstract}
A single-electron transistor (SET) in a magnetic field irradiated
with microwaves is studied theoretically in non-equilibrium Kondo
regime. The two fold effect of frequency--$\Omega$--microwaves is
considered as follows: the oscillations in the voltage with frequency
$\Omega$ and in the coupling parameters with frequency $\Omega/p$
($p\in\mathbb{N}$). We describe the system by the Kondo model at
a specific point in the Toulouse limit. A non-perturbative technique
is proposed, namely, the non-equilibrium Green's functions and physical
observables are averaged over a period of $2\pi p/\Omega$. When the
microwave irradiation is considered affecting only the voltage, one
sees the Kondo satellites as stated in the previous studies. Moreover,
the features of the differential conductance and magnetic susceptibility
of a SET become richer when the Kondo couplings are considered oscillating
on time. We obtain the satellite peak splitting. It explains the possibilities
one can find in experimental results that the distance between peaks,
which appear in the differential conductance -- magnetic amplitude
characteristics $G\left(H\right)$ or in the differential conductance
-- dc voltage characteristics $G\left(V_{dc}\right)$, can be smaller
than $\hbar\Omega$.
\end{abstract}

\pacs{72.10.Fk, 72.15.Qm, 73.40.Gk }

\maketitle

\section{Introduction}

The Kondo effect, stemming from a macroscopic quantum coherent coupling
between a localized magnetic moment and a Fermi sea of electrons,
has been used to explain successfully many extraordinary properties
of bulk dilute magnetic alloy systems \cite{JKondo,Kondo_book} for
a long time. Based on the developments in mesoscopic physics and nano-technology,
it has been proposed to study the systems of reduced dimensions and
the single impurity problems. The latter is also represented by the
very rapidly developing fields of quantum dots (QDs), where Kondo-like
impurities are manufactured \cite{KondoQD,KondoQDexperiments,reviewKondoQD}.
The local magnetic moment is screened by hybridization with the delocalized
electron-spins, leading to the formation of a bound spin-singlet state.
A many-body resonance in the density of states (DOS) appears at the
Fermi energy, which strongly influences the conductance of the system
at low temperatures. Above the Kondo temperature, thermal fluctuations
destroy the coherence, resulting in a non-monotonic dependence of
the conductance on temperature. Under the Kondo temperature, due to
the Kondo correlations, the system becomes dynamic and can be controlled
experimentally by tuning external parameters such as source-drain
voltage, gate voltage, and magnetic field \cite{KondoQDexperiments}. 

Since the Kondo effect is a typical example of a strongly correlated
system, it is of particular interest in the field of non-equilibrium
phenomena \cite{KondoQDnoneq1,KondoQDnoneq2}. As our understanding
of the Kondo effect for a time-independent bias has increased \cite{KondoQDnoneq1,KondoQDnoneq2},
the theoretical effort has turned to time-varying fields \cite{Hettler_Schoeller,TKNg,SchillerHershfield,Lopez,Avishai,Kaminski,Nordlander}.
The ac field produces satellite peaks in the dependence of the differential
conductance on the dc bias, as a consequence of the fact that the
microwave field gives rise to effective changes in the DOS versus
energy which are dramatically illustrated in the tunneling current
\cite{Tien_Gordon}. However, this interesting result could not be
observed experimentally until 2004 by Kastner's group \cite{sideband_exp}.
They measured the differential conductance of a SET irradiated with
high frequency microwaves. The results show that when the photon energy
is somewhat greater than the Kondo temperature and the microwave voltage
is carefully chosen, single-photon satellite peaks in the differential
conductance can be observed. Very recently, Kogan's group has investigated
the differential conductance as a function of magnetic field for different
high frequencies of microwaves \cite{Experiment}. The non-trivial
dynamics that emerges when external parameters compete with the Kondo
correlations has attracted our attention to study the Kondo effect
in a SET irradiated with microwaves. Due to the complexity of out-of-equilibrium
many-body systems, it is instructive to investigate those cases where
the exact non-perturbative solutions are accessible. One particular
interesting case is the Kondo model in the Toulouse limit, which has
been solved exactly in equilibrium \cite{EmeryKevilson} as well as
out of equilibrium (SH theory) \cite{SchillerHershfield}. A question
arising from the SH theory \cite{SchillerHershfield}, which is related
to the experiment, is what happens if the Kondo coupling parameters
are also affected by microwave irradiation. We can have different
scenarios in different experiments: i, only the voltage oscillates;
ii, only the Kondo couplings oscillate; iii, both of them oscillate
\cite{discussion}. Moreover, the multi-effect of microwave irradiation
can induce the fact that the voltage oscillates with frequency $\Omega$,
the Kondo couplings oscillate with frequency $\Omega_{1}=\Omega/p$,
$p\in\mathbb{N}$ in an appropriate experimental set-up. We predict
that $p$ is a frequency-dependent systematic parameter.

In this paper we investigate the response of a SET to a simultaneous
application of a magnetic field and a high frequency microwave irradiation.
We consider Kondo couplings oscillating with a frequency integer times
smaller than the frequency of microwaves. We solve the problem non-perturbatively
by averaging non-equilibrium Green's functions in the period of the
slow oscillation. The differential conductance and the magnetic susceptibility
are computed. In these results, the Kondo satellites appear due to
the oscillation of the source-drain voltage. These satellites are
split when the oscillation of Kondo couplings is taken into account.
The peak splitting distance depends on the difference between the
input frequency and the frequency of the Kondo couplings, which is
assumed to be a natural number $p$. 

This paper is organized as follows. In section \ref{sec:Kondo-model},
the time-dependent Kondo model in the Toulouse limit with time dependent
Kondo couplings $J_{\lambda}^{\alpha\beta}(t)$ is introduced and
mapped onto a quadratic effective Hamiltonian. The averaged non-equilibrium
Green's function method used to calculate the differential conductance
and the magnetic susceptibility is presented in section \ref{sec:Average-Green's-functions}.
The results for the differential conductance and the magnetic susceptibility
are presented in section \ref{sec:average-physical-observables}.
We conclude our results in section \ref{sec:Conclusions}.

\section{Time dependent Kondo model \label{sec:Kondo-model}}

\subsection{Model and proposal for time-dependent Kondo couplings}

The system we study is a QD attached to two leads by high resistance
junctions, so that the charge of the dot is quantized. The Kondo effect
emerges in a QD occupied by an odd number of electrons at temperatures
below the mean level spacing in the dot. Under such conditions, the
system can be described by a single-impurity Anderson model with the
level spacing playing the role of cutoff in it \cite{Anderson_model}.
Moreover, we consider the dot in the Kondo regime: both ionization
and electron addition energy are much bigger than the tunneling rates:
$\left\{ E_{d},\, U-E_{d}\right\} \gg\Gamma_{L/R}$, and the applied
fields do not drive the dot out of this regime: $\left\{ eV_{dc},\, eV_{dot},\, eV_{ac},\:\mu_{B}g_{i}B\right\} <\left\{ E_{d},\, U-E_{d}\right\} $.
Hence, the Anderson model can be mapped onto a two lead Kondo model
through a time dependent Schrieffer-Wolff transformation \cite{Avishai,Kaminski}.

In our proposed Kondo model, the two non-interacting leads of spin-1/2
electrons, labeled right (R) and left (L), each subjects to a separate
time-dependent voltage, interact via an exchange coupling with a spin-1/2
impurity moment $\overrightarrow{\tau}$ placed in between them. The
conduction electrons in each lead, that couple to the impurity, are
described by one dimensional fields $\psi_{\alpha\sigma}^{\dagger}\left(x\right)$,
where $\alpha=L,R$, and $\sigma=\uparrow,\downarrow$ are lead and
spin indices, respectively. The interaction $\overrightarrow{J}^{\alpha\beta}$
of conduction electrons in the leads and the impurity is local in
space at $x=0$ and involves the conduction-electron spin densities,
which are written as $\left(\overrightarrow{s}\right)_{\alpha,\beta}=\frac{1}{2}\sum_{\sigma,\sigma^{\prime}}\psi_{\alpha\sigma}^{\dagger}\left(0\right)\left(\overrightarrow{\sigma}\right)_{\sigma,\sigma^{\prime}}\psi_{\beta\sigma^{\prime}}\left(0\right)$.
Thus, the time-dependent Kondo Hamiltonian describing a SET in a magnetic
field and irradiated with microwaves is

\begin{align}
\mathcal{H} & =iv_{F}\sum_{\alpha=L,R;\sigma=\uparrow,\downarrow}\int_{-\infty}^{\infty}dx\psi_{\alpha\sigma}^{\dagger}\frac{\partial}{\partial x}\psi_{\alpha\sigma}\nonumber \\
 & +\frac{V\left(t\right)}{2}\sum_{\sigma=\uparrow,\downarrow}\int_{-\infty}^{\infty}dx\left[\psi_{L\sigma}^{\dagger}\psi_{L\sigma}-\psi_{R\sigma}^{\dagger}\psi_{R\sigma}\right]\nonumber \\
 & +\sum_{\alpha,\beta=L,R}\sum_{\lambda=x,y,z}J_{\lambda}^{\alpha\beta}\left(t\right)s_{\alpha\beta}^{\lambda}\tau^{\lambda}-H\tau^{z}\:.\label{eq:Kondo0}
\end{align}
The system is driven out of equilibrium by applying a voltage bias
across the junction, $V\left(t\right)=V_{dc}+V_{ac}\cos\left(\Omega t\right)$.
The dc part $V_{dc}$ fixes a chemical-potential difference between
the two Fermi seas. The ac part $V_{ac}\cos\left(\Omega t\right)$
fluctuates these two chemical-potentials as an oscillation on time,
allowing us to average the Green's functions, which will be shown
in detail in the next section. We notice that we adopt atomic units
with $\hbar=k_{B}=e=\mu_{B}g_{i}=1$ in this paper.

In general, due to the microwave irradiation, the couplings are split
into two parts: the time-independent part and the time-dependent part,
which oscillates on time, as
\begin{equation}
J_{\lambda}^{\alpha\beta}\left(t\right)=J_{\lambda0}^{\alpha\beta}+J_{\lambda1}^{\alpha\beta}\cos\left(\Omega_{1}t+\phi_{\alpha\beta}\right)\:,\label{eq:condition2}
\end{equation}
 $\alpha,\,\beta=L,\, R$ imply the left and right leads. The oscillation
frequency $\Omega_{1}$ of couplings is different from the microwave
frequency $\Omega$. In order to enter the exact solvable situation,
we assume \cite{SchillerHershfield,EmeryKevilson,SchillerHershfield-PRB}
\begin{equation}
J_{z}^{LR}=J_{z}^{RL}=0,\; J_{z}^{LL}=J_{z}^{RR}=J_{z}=2\pi v_{F}\:.\label{eq:condition1}
\end{equation}
If the SET is in equilibrium or if a dc voltage is applied, the coupling
parameters are constant and equal to $J_{\perp0}^{\alpha\beta}$.
In equilibrium case, from the scaling equations to lowest order in
the couplings, we find that the constraints in equation (\ref{eq:condition1})
remain stable upon scaling if either 
\begin{equation}
J_{\perp0}^{LL}=J_{\perp0}^{RR}\:,\; J_{\perp0}^{RL}=0\:,\label{eq:assumption}
\end{equation}
or 
\begin{equation}
J_{\perp0}^{LL}=-J_{\perp0}^{RR}\:.
\end{equation}
As stated in Ref. \cite{SchillerHershfield-PRB}, these two cases
correspond to two distinct isotropic two-channel Kondo model. We notice
that in the constraint (\ref{eq:assumption}), the two leads are decoupled,
the channels are just the right and left leads, which carry no current.
When the system is irradiated with microwaves, both the voltage bias
and the perpendicular couplings are proposed oscillating on time.
The model in this paper is anisotropic in real space. It cannot be
obtained by the direct mapping from the Anderson model to the Kondo
model. However, the experimentalists can apply the relations between
the Kondo couplings and the tunneling rates, which are stated in Refs.
\cite{Avishai,Kaminski}, in order to compare their experimental results
to the results in our theory.

For simplicity, we propose the time dependent perpendicular coupling
parameters as 
\begin{eqnarray}
J_{\perp}^{RL}(t) & = & J_{\perp}^{LR}(t)=J_{\perp}^{RL}\cos\left(\Omega_{1}t\right)\,,\nonumber \\
J_{\perp}^{LL}(t) & = & J_{\perp0}^{LL}+J_{\perp}^{LL}\cos\left(\Omega_{1}t\right)\,,\nonumber \\
J_{\perp}^{RR}(t) & = & J_{\perp0}^{LL}+J_{\perp}^{RR}\cos\left(\Omega_{1}t\right)\,.\label{eq:parameters}
\end{eqnarray}
This proposal can be justified qualitatively as follows. The oscillation
of $J(t)$ comes from the oscillations of voltages such as source-drain
voltage and gate voltage. The voltage biases oscillate with the same
frequency $\Omega$, as the frequency of microwaves. The Kondo couplings
thus oscillate with frequency $\Omega$. However, experimentalists
can think about some experimental set-ups, in which one can control
the frequencies of different voltages or the phases between voltage
biases oscillate on time. The goal of this work is to have the Kondo
couplings $J(t)$ oscillating on time with frequency $\Omega_{1}\neq\Omega$.
For simplicity, we assume either $\Omega_{1}=p\Omega$ or $\Omega_{1}=\Omega/p$,
$p\in\mathbb{N}$. Since the $\Omega_{1}=\Omega/p$ assumption is
predicted to give the results of satellite splitting, it is chosen
to be studied first. The $\Omega_{1}=p\Omega$ assumption will be
represented later. 

Besides, one can investigate the time dependence of Kondo couplings
in a general form as shown in equation (\ref{eq:condition2}). However,
the goal of this paper is to investigate the effects of the microwave
irradiation. We thus consider the Kondo couplings as shown in equations
(\ref{eq:parameters}). Since we consider the time average of observables,
we can neglect the phase difference among the coupling parameters.
In general, microwave irradiation changes the transverse couplings
$J_{\perp}^{RL}(t)$, thus changes the channel isotropy. We find that
a QD either in a magnetic field \cite{Kondo_channel}, or irradiated
with microwaves can be driven out of the isotropic two channel Kondo
limit.

\subsection{Mapping onto a solvable model}

The procedure to arrive at a solvable model is given in the following.
We first bosonize the Hamiltonian (\ref{eq:Kondo0}) with notice that
we have considered Klein factors and assumed the length was infinity:
\begin{equation}
\psi_{\alpha\sigma}(x)=\frac{F_{\alpha\sigma}}{\sqrt{2\pi a}}e^{-i\Phi_{\alpha\sigma}(x)}\;,\label{fermion}
\end{equation}
 where $\Phi_{\alpha\sigma}(x)=\sqrt{\pi}\!\left[\intop_{-\infty}^{x}\!\!\!\! dx^{\prime}\Pi_{\alpha\sigma}(x^{\prime})-\phi_{\alpha\sigma}(x)\right]$.
Here $\phi_{\alpha\sigma}(x)$ are Bose fields and $\Pi_{\alpha\sigma}(x)$
are their conjugate momenta, satisfying commutation relations \cite{bosonization}.
The next step is introducing new bosonic fields: charge $\phi_{c}(x)$,
pseudo-spin $\phi_{s}(x)$, flavor $\phi_{f}(x)$, and pseudo-flavor
$\phi_{sf}(x)$ as $\phi_{c}(x)=\left(\sum_{\alpha,\sigma}\phi_{\alpha\sigma}\right)/2$,
$\phi_{s}(x)=\left(\sum_{\alpha,\sigma}\sigma_{\sigma\sigma}^{z}\phi_{\alpha\sigma}\right)/2$,
$\phi_{f}(x)=\left(\sum_{\alpha,\sigma}\sigma_{\alpha\alpha}^{z}\phi_{\alpha\sigma}\right)/2$,
$\phi_{sf}(x)=\left(\sum_{\alpha,\sigma}\sigma_{\alpha\alpha}^{z}\sigma_{\sigma\sigma}^{z}\phi_{\alpha\sigma}\right)/2$,
and also for $\Pi_{\nu},\;\Phi_{\nu},\; N_{\nu}$, $\nu=c,\, s,\, f,\, sf$.
We then perform the transformation of the Hamiltonian $\mathcal{UHU}^{-1}$with
$\mathcal{U}=\exp\left[-i\tau^{z}\Phi_{s}(0)\right]$. Four more Klein
factors $F_{\nu}$ are introduced, which satisfy $\left[F_{\nu},N_{\nu^{\prime}}\right]=\delta_{\nu\nu^{\prime}}F_{\nu}$
and relate to the old ones as $F_{L\downarrow}^{\dagger}F_{L\uparrow}=F_{s}F_{sf}$,
$F_{R\downarrow}^{\dagger}F_{R\uparrow}=F_{s}F_{sf}^{\dagger}$, $F_{R\downarrow}^{\dagger}F_{L\uparrow}=F_{f}F_{s}$,
$F_{R\uparrow}^{\dagger}F_{L\downarrow}=F_{f}F_{s}^{\dagger}$. The
new impurity fermion is thus represented as $d^{\dagger}=F_{s}\tau^{+}$,
$d=F_{s}^{\dagger}\tau^{-}$, $\tau^{z}=d^{\dagger}d-1/2$. We now
re-fermionize these bosonic fields as 
\begin{equation}
\Psi_{m}(x)=\frac{1}{\sqrt{2\pi a}}F_{m}e^{-i\Phi_{m}(x)}
\end{equation}
to re-write the Hamiltionian in which all the charge, spin, flavor,
and spin flavor degrees of freedom are separated 
\begin{align}
 & \!\!\!\mathcal{H}^{\prime}=iv_{F}\sum_{\nu=c,s,f,sf}\int_{-\infty}^{\infty}\!\!\!\!\!\!\! dx\psi_{\nu}^{\dagger}(x)\partial_{x}\psi_{\nu}(x)\nonumber \\
 & \!\!\!+\left[H-\left(J_{z}-2\pi v_{F}\right):\psi_{s}^{\dagger}(0)\psi_{s}(0):\right]\left(d^{\dagger}d-1/2\right)\nonumber \\
 & \!\!\!+\int_{-\infty}^{\infty}\!\!\!\!\!\!\! dx\,\left[V_{dc}+V_{ac}\cos(\Omega t)\right]\psi_{f}^{\dagger}(x)\psi_{f}(x)\nonumber \\
 & \!\!\!+\frac{J_{t}(t)}{2\sqrt{2\pi a}}\!\!\left[\Psi_{f}^{\dagger}(0)-\Psi_{f}(0)\right]\!\!\left(d+d^{\dagger}\right)\nonumber \\
 & \!\!\!+\frac{J_{a}(t)}{2\sqrt{2\pi a}}\!\!\left[\Psi_{sf}^{\dagger}(0)-\Psi_{sf}(0)\right]\!\!\!\left(d+d^{\dagger}\right)\nonumber \\
 & \!\!\!+\frac{J_{s}(t)}{2\sqrt{2\pi a}}\!\!\left[\Psi_{sf}^{\dagger}(0)+\Psi_{sf}(0)\right]\!\!\!\left(d^{\dagger}-d\right)\:,\label{eq:Kondo1}
\end{align}
where $:\psi_{s}^{\dagger}(0)\psi_{s}(0):$ means normal ordering
with respect to the unperturbed $\psi_{s}$ Fermi sea, and we have
defined $J_{t}(t)=J_{\perp}^{LR}(t)=J_{\perp}^{RL}\cos\left(\Omega_{1}t\right)$,
$J_{s}(t)=\left(J_{\perp0}^{LL}+J_{\perp0}^{RR}\right)/2+\left(J_{\perp}^{LL}+J_{\perp}^{RR}\right)\cos\left(\Omega_{1}t\right)/2$,
$J_{a}(t)=\left(J_{\perp}^{LL}-J_{\perp}^{RR}\right)\cos\left(\Omega_{1}t\right)/2$.
From our assumption in equation (\ref{eq:condition1}), $J_{z}=2\pi v_{F}$,
we are in the Emery-Kivelson line \cite{EmeryKevilson}, the Hamiltonian
(\ref{eq:Kondo1}) reduces to quadratic form, which is exactly solvable
in equilibrium \cite{EmeryKevilson} and in the SH theory \cite{SchillerHershfield,SchillerHershfield-PRB}.
The charge and spin sectors are decoupled from the local flavor and
spin flavor, reduced to a collection of uncoupled harmonic oscillators.
Therefore, they will be omitted from now on. Finally, by introducing
Majorana fermions $\hat{a}=\left(d^{\dagger}+d\right)/\sqrt{2}$ and
$\hat{b}=\left(d^{\dagger}-d\right)/i\sqrt{2}$, one arrives at the
following Hamiltonian 
\begin{align}
\mathcal{H} & =\sum_{k}\left[\varepsilon_{fk}(t)c_{fk}^{\dagger}c_{fk}+\varepsilon_{sfk}c_{sfk}^{\dagger}c_{sfk}\right.\nonumber \\
 & \left.-iH\hat{a}\hat{b}+\frac{J_{t}(t)}{2\sqrt{\pi a}}(c_{fk}^{\dagger}-c_{fk})\hat{a}\right.\nonumber \\
 & \left.+\frac{iJ_{s}(t)}{2\sqrt{\pi a}}(c_{sfk}^{\dagger}+c_{sfk})\hat{b}+\frac{J_{a}(t)}{2\sqrt{\pi a}}(c_{sfk}^{\dagger}-c_{sfk})\hat{a}\right]\,,\label{eq:hamiltonian}
\end{align}
where $\varepsilon_{fk}(t)=2\pi v_{F}k+e[V_{dc}+V_{ac}\cos\left(\Omega t\right)]$,
$\varepsilon_{sfk}=2\pi v_{F}k$. Here after we will work with Hamiltonian
(\ref{eq:hamiltonian}) to calculate physical observables.

\section{Average Green's functions \label{sec:Average-Green's-functions}}

The Keldysh Green's function technique is well known as an efficient
method to solve non-equilibrium problems \cite{reviewKeldyshGreenfunction}.
This technique has been applied to express the fully nonlinear, time
dependent current through interacting and non-interacting resonant
tunneling systems \cite{KeldyshQD}. However, the complete result
only obtained for the case level-width function (or tunneling amplitude)
is time independent. Similarly, the Hamiltonian (\ref{eq:hamiltonian})
with constant couplings has been solved exactly \cite{SchillerHershfield}.
Later, the steady state in the Kondo model in the Toulouse limit,
in which the couplings are periodically switched on and off, has been
investigated by analyzing exact analytical results for the local spin
dynamics at zero temperature \cite{Kehrein}. The exact solution for
the time dependent Hamiltonian (\ref{eq:hamiltonian}) still remains.
In this section, we present our average Keldysh non-equilibrium Green's
function method which gives good results in the high frequency regime.

\subsection{Non-interacting Green's functions of flavor fermions}

Because the first effect of microwave irradiation induces the oscillation
in the voltage bias as shown in the Hamiltonian (\ref{eq:hamiltonian}),
we first discuss the non-interacting Green's functions of the flavor
fermion, which is defined as 
\begin{equation}
g_{fk}\left(t,t^{\prime}\right)=-i\left\langle T_{K}\left\{ c_{fk}\left(t\right)c_{fk}^{\dagger}\left(t^{\prime}\right)\right\} \right\rangle \;.
\end{equation}
The flavor fermion is considered as in the two dimensional fermion
sea. The retarded, advanced, and Keldysh components are 

\begin{align}
g_{fk}^{R,A}\left(t,t^{\prime}\right) & =\mp i\Theta\left(\pm t\mp t^{\prime}\right)e^{-i2\pi v_{F}k\left(t-t^{\prime}\right)}\nonumber \\
 & \times e^{-i\frac{V_{ac}}{\Omega}\left[\sin\left(\Omega t\right)-\sin\left(\Omega t^{\prime}\right)\right]}\;,\nonumber \\
g_{fk}^{K}\left(t,t^{\prime}\right) & =i\left[2f\left(2\pi v_{F}k+V_{dc}\right)-1\right]e^{-i2\pi v_{F}k\left(t-t^{\prime}\right)}\nonumber \\
 & \times e^{-i\frac{V_{ac}}{\Omega}\left[\sin\left(\Omega t\right)-\sin\left(\Omega t^{\prime}\right)\right]}\;,
\end{align}
where $f$ is the Fermi distribution function in the flavor lead.
While the chemical potential of the spin flavor lead is chosen as
the reference, the chemical potential of the flavor lead oscillates
on time with an amplitude $V_{ac}$ around the fixed chemical potential
$V_{dc}$. Using the identity $\exp\left[x\left(a-1/a\right)/2\right]=\sum_{n=-\infty}^{\infty}a^{n}J_{n}\left(x\right)$,
where $J_{n}\left(x\right)$ are the integer Bessel functions of the
first kind, and changing variables
\begin{eqnarray}
\tau & = & t-t^{\prime}\:,\nonumber \\
T & = & \frac{t+t^{\prime}}{2}\;,\label{eq:change_variables}
\end{eqnarray}
we find 
\begin{align}
g_{fk}^{R,A}(\tau,T) & =\mp i\Theta(\pm\tau)\sum_{m,n=-\infty}^{\infty}\!\!\!\!\!\!\! J_{n}\left(\frac{V_{ac}}{\Omega}\right)J_{m}\left(\frac{V_{ac}}{\Omega}\right)\nonumber \\
 & \times e^{-i\left[\varepsilon_{k}+\left(n+m\right)\Omega/2\right]\tau}e^{-i\Omega\left(n-m\right)T}\:.
\end{align}

Because the flavor fermion chemical potential oscillates on time with
frequency $\Omega$, one can take the average of the non-interacting
Green's function over time $T\gg\Omega^{-1}$\cite{Mitra}. The Fourier
transform regarding the time difference $\tau$ of the average Green's
function is 
\begin{align}
 & \overline{g_{f}^{R,A}(\omega)}=\sum_{k}\overline{g_{fk}^{R,A}(\omega)}\nonumber \\
 & =\sum_{m,n=-\infty}^{\infty}\!\!\!\!\!\!\! J_{n}^{2}\left(\frac{V_{ac}}{\Omega}\right)\sum_{k}\frac{1}{\omega-2\pi v_{F}k-n\Omega}\;.
\end{align}

The DOS of the flavor fermion is \cite{Tien_Gordon} 
\begin{align}
\overline{\rho_{f}\left(\omega\right)} & =\sum_{n}J_{n}^{2}\left(\frac{V_{ac}}{\Omega}\right)\rho\left(\omega-n\Omega\right)\nonumber \\
 & =\sum_{n}J_{n}^{2}\left(\frac{V_{ac}}{\Omega}\right)\frac{1}{2\pi v_{F}}=\frac{1}{2\pi v_{F}}\;.
\end{align}
We have the second line because we have applied the linearization
for the energy spectrum around the Fermi level \cite{bosonization}.
The retarded and advanced components thus remain as those in equilibrium
\begin{equation}
\overline{g_{f}^{R,A}(\omega)}=\mp\frac{i}{2v_{F}}\;.\label{eq:-1-2}
\end{equation}
However, the effect of ac voltage $V_{ac}\left(t\right)$ modifies
the Keldysh component $g_{fk}^{K}\left(t,t^{\prime}\right)$ as

\begin{align}
 & \overline{g_{f}^{K}(\omega)}=\sum_{k}\overline{g_{fk}^{K}(\omega)}\nonumber \\
 & =\frac{i}{v_{F}}\sum_{n}J_{n}^{2}\left(\frac{V_{ac}}{\Omega}\right)\left[2f\left(\omega-n\Omega+V_{dc}\right)-1\right]\;,
\end{align}
with $\overline{g_{f}^{K}(\omega)}$ is the average of the Keldysh
component in the Fourier space corresponding the time difference $\tau$.

We then calculate the combined non-interacting Green's functions of
the flavor fermion. We show here one example 

\begin{align}
\!\!\! m_{fk}\!\left(t,t^{\prime}\right) & \!=\!-i\left\langle T_{K}\!\!\left[c_{fk}^{\dagger}\!\left(t\right)\!+c_{fk}\left(t\right)\!\right]\!\left[c_{fk}^{\dagger}\!\left(t^{\prime}\right)\!-c_{fk}\!\left(t^{\prime}\right)\!\right]\right\rangle \:.
\end{align}
Its retarded, advanced, and Keldysh components are $\overline{m_{f}^{R,A}(\omega)}=0$,
\begin{align}
 & \!\!\!\!\!\!\overline{m_{f}^{K}(\omega)}=\left(2i/v_{F}\right)\sum_{n}J_{n}^{2}\left(\frac{V_{ac}}{\Omega}\right)\nonumber \\
 & \!\!\!\!\!\!\times\left\{ f\left(\omega-n\Omega+V_{dc}\right)-f\left(\omega+n\Omega-V_{dc}\right)\right\} \,.
\end{align}

About the spin flavor fermion, its chemical potential is chosen as
the reference, and no oscillation voltage is applied on the spin flavor
lead, it is considered in equilibrium.

\subsection{Interacting Majorana Green's functions:}

In order to compute observables, it is necessary to calculate advanced,
retarded, and Keldysh components of Majorana Green's functions, which
are defined as
\begin{eqnarray}
\!\!\!\!\!\!\!\!\!\! G_{\alpha\beta}\left(t,t^{\prime}\right) & = & -i\left\langle T_{K}\left\{ \alpha\left(t\right)\beta\left(t^{\prime}\right)\right\} \right\rangle \:,\;\alpha,\beta=\hat{a},\hat{b}\:.\label{eq:int_Greens}
\end{eqnarray}
We show here the results of the Green's functions $G_{aa}^{A}(t,t^{\prime})$
and $G_{ba}^{K}(t,t^{\prime})$, which are used to calculate the differential
conductance and the magnetic susceptibility.

\subsubsection{Advanced Green's function $G_{aa}^{A}(t,t^{\prime})$:}

We write the Green's functions in terms of the interaction-picture
operators by invoking the S matrix, we have the advance Green's functions
as follows

\begin{align}
G_{ba}^{A}(t,t^{\prime}) & =iH\int_{-\infty}^{\infty}\!\!\!\!\! dt_{1}g_{bb}^{A}(t-t_{1})G_{aa}^{A}(t_{1},t^{\prime})\nonumber \\
 & \!\!\!\!\!+\frac{i}{4\pi av_{F}}\int_{-\infty}^{\infty}\!\!\!\!\! dt_{1}J_{s}^{2}\left(t_{1}\right)g_{bb}^{A}(t-t_{1})G_{ba}^{A}(t_{1},t^{\prime})\:,\nonumber \\
G_{aa}^{A}(t,t^{\prime}) & =g_{aa}^{A}(t,t^{\prime})-iH\int_{-\infty}^{\infty}\!\!\!\!\! dt_{1}g_{aa}^{A}(t-t_{1})G_{ba}^{A}(t_{1},t^{\prime})\nonumber \\
 & \!\!\!\!\!\!\!\!\!\!\!\!\!\!\!\!\!\!\!\!\!\!\!\!\!\!\!\!\!\!+\frac{i}{4\pi av_{F}}\int_{-\infty}^{\infty}\!\!\!\!\! dt_{1}\left[J_{t}^{2}\left(t_{1}\right)+J_{a}^{2}\left(t_{1}\right)\right]g_{aa}^{A}(t-t_{1})G_{aa}^{A}(t_{1},t^{\prime})\;.\label{eq:Gaa-2}
\end{align}

With the assumption of coupling parameters $J_{i}\left(t\right)$
from (\ref{eq:hamiltonian}), we have 
\begin{align}
 & G_{ba}^{A}(t,t^{\prime})=-H\int_{-\infty}^{\infty}\!\!\!\!\! dt_{1}\Theta(t_{1}-t)G_{aa}^{A}(t_{1},t^{\prime})\nonumber \\
 & -\int_{-\infty}^{\infty}\!\!\!\!\! dt_{1}\Gamma_{s}\left(t_{1}\right)\Theta(t_{1}-t)G_{ba}^{A}(t_{1},t^{\prime})\;,\nonumber \\
 & G_{aa}^{A}(t,t^{\prime})=i\Theta(t^{\prime}-t)+H\int_{-\infty}^{\infty}\!\!\!\!\! dt_{1}\Theta(t_{1}-t)G_{ba}^{A}(t_{1},t^{\prime})\nonumber \\
 & -\int_{-\infty}^{\infty}\!\!\!\!\! dt_{1}\Gamma_{at}\left(t_{1}\right)\Theta(t_{1}-t)G_{aa}^{A}(t_{1},t^{\prime})\;,\label{eq:Gaa_equations}
\end{align}
with 
\begin{align}
\Gamma_{s}\left(t\right) & =\Gamma_{s0}+2\sqrt{\Gamma_{s0}\Gamma_{s}}\cos\left(\Omega_{1}t\right)+\Gamma_{s}\cos^{2}\left(\Omega_{1}t\right)\:,\nonumber \\
\Gamma_{at}\left(t\right) & =\Gamma_{at}\cos^{2}\left(\Omega_{1}t\right)\:,\label{eq:energies}
\end{align}
where we have defined $\Gamma_{t}=\left(J_{\perp}^{RL}\right)^{2}/4\pi av_{F}$,
$\Gamma_{a}=\left(J_{\perp}^{LL}-J_{\perp}^{RR}\right)^{2}/16\pi av_{F}$,
$\Gamma_{at}=\Gamma_{t}+\Gamma_{a}$, $\Gamma_{s0}=\left(J_{\perp0}^{LL}+J_{\perp0}^{RR}\right)^{2}/16\pi av_{F}$,
$\Gamma_{s}=\left(J_{\perp}^{LL}+J_{\perp}^{RR}\right)^{2}/16\pi av_{F}$.
Equations (\ref{eq:Gaa_equations}) induces the SH theory limit when
all couplings are time independent. One can easily take Fourier transform
of equations (\ref{eq:Gaa_equations}), then re-obtain the exact formulas
of the advanced Majorana Green's functions, as shown in the formula
(4.9) in Ref. \cite{SchillerHershfield-PRB}.

We need to solve the integral equations (\ref{eq:Gaa_equations})
and find the Green's function $G_{aa}^{A}(t,t^{\prime})$. We first
change the equations (\ref{eq:Gaa_equations}) to differential equations
by taking the derivative with respect to time $t$. We have 
\begin{align}
 & \!\!\!\!\left[\partial_{t}-\Gamma_{s}\left(t\right)\right]G_{ba}^{A}\left(t,t^{\prime}\right)=HG_{aa}^{A}\left(t,t^{\prime}\right)\,,\nonumber \\
 & \!\!\!\!\left[\partial_{t}-\Gamma_{at}\left(t\right)\right]G_{aa}^{A}\left(t,t^{\prime}\right)+HG_{ba}^{A}\left(t,t^{\prime}\right)=-i\delta\left(t-t^{\prime}\right)\,.\label{eq:solve1}
\end{align}
We define the non-interacting Green's functions $g_{at/s}^{A}\left(t,t^{\prime}\right)$
as \cite{Green's funtion} 
\begin{align}
\left[\partial_{t}-\Gamma_{at/s}\left(t\right)\right]g_{at/s}^{A}\left(t,t^{\prime}\right) & =\delta\left(t-t^{\prime}\right)\,,\label{eq:green0}
\end{align}
so
\begin{align}
g_{at}^{A}\left(t,t^{\prime}\right) & =-\Theta(t^{\prime}-t)\exp\left[\frac{\Gamma_{at}}{2}\left(t-t^{\prime}\right)\right]\nonumber \\
 & \times\exp\left[\frac{\Gamma_{at}}{4\Omega_{1}}\left[\sin\left(2\Omega_{1}t\right)-\sin\left(2\Omega_{1}t^{\prime}\right)\right]\right]\,,\nonumber \\
g_{s}^{A}\left(t,t^{\prime}\right) & =-\Theta(t^{\prime}-t)\exp\left[\left(\Gamma_{s0}+\frac{\Gamma_{s}}{2}\right)\left(t-t^{\prime}\right)\right]\nonumber \\
 & \times\exp\left[\frac{2\sqrt{\Gamma_{s0}\Gamma_{s}}}{\Omega_{1}}\left[\sin\left(\Omega_{1}t\right)-\sin\left(\Omega_{1}t^{\prime}\right)\right]\right]\nonumber \\
 & \times\exp\left[\frac{\Gamma_{s}}{4\Omega_{1}}\left[\sin\left(2\Omega_{1}t\right)-\sin\left(2\Omega_{1}t^{\prime}\right)\right]\right]\label{eq:gA0}
\end{align}

From equations $\left(\ref{eq:solve1}\right)$ and $\left(\ref{eq:green0}\right)$,
we have 
\begin{align}
 & \!\!\!\!\int_{-\infty}^{\infty}\!\!\!\!\!\!\!\! dt_{1}\!\left[\!\left[g_{at}^{A}\left(t,t_{1}\right)\right]^{-1}\!\!\!+H^{2}g_{s}^{A}\left(t,t_{1}\right)\!\right]\! G_{aa}^{A}\left(t_{1},t^{\prime}\right)\!=\!-i\delta\left(t-t^{\prime}\right)\,.\label{eq:solve2}
\end{align}

\subsubsection{Keldysh Green's function $G_{ba}^{K}(t,t^{\prime})$: }

In the same way as we did with the advanced Green's functions, we
write the Keldysh Green's functions by invoking the S-matrix as
\begin{align}
 & \!\!\!\!\! G_{ba}^{K}(t,t^{\prime})=\int_{-\infty}^{\infty}\!\!\!\!\! dt_{1}H\Theta(t-t_{1})G_{aa}^{K}(t_{1},t^{\prime})\nonumber \\
 & \!\!\!\!\!-\int_{-\infty}^{\infty}\!\!\!\!\! dt_{1}\Gamma_{s}\left(t_{1}\right)\Theta(t-t_{1})G_{ba}^{K}(t_{1},t^{\prime})+H\Upsilon_{s}\left(t,t^{\prime}\right)\:,
\end{align}

\begin{align}
 & \!\!\!\!\! G_{aa}^{K}(t,t^{\prime})=\!-\!\int_{-\infty}^{\infty}\!\!\!\!\! dt_{1}H\Theta(t-t_{1})G_{ba}^{K}(t_{1},t^{\prime})\nonumber \\
 & \!\!\!\!\!-\int_{-\infty}^{\infty}\!\!\!\!\! dt_{1}\Gamma_{at}\left(t_{1}\right)\Theta(t-t_{1})G_{aa}^{K}(t_{1},t^{\prime})+\Upsilon_{at}\left(t,t^{\prime}\right)\:,
\end{align}
where we have defined the $\Upsilon_{at}\left(t,t^{\prime}\right)$,
$\Upsilon_{s}\left(t,t^{\prime}\right)$ as 
\begin{align}
\Upsilon_{s}\left(t,t^{\prime}\right) & =\int_{-\infty}^{\infty}\!\!\!\!\!\! dt_{1}dt_{2}\frac{J_{s}\left(t_{1}\right)J_{s}\left(t_{2}\right)}{4\pi^{2}av_{F}}\Theta(t-t_{1})\int_{-\infty}^{\infty}\!\!\!\!\! d\omega\nonumber \\
 & \!\!\!\!\!\!\!\!\!\!\!\!\!\!\!\!\!\!\!\!\!\!\!\!\!\!\!\!\!\!\!\!\!\!\!\!\times\left[2f\left(\omega\right)-1\right]e^{-i\omega(t_{1}-t_{2})}\int_{-\infty}^{\infty}\!\!\!\!\! dt_{3}g_{s}^{A}\left(t_{2},t_{3}\right)G_{aa}^{A}(t_{3},t^{\prime})\:,
\end{align}
\begin{align}
 & \Upsilon_{at}\left(t,t^{\prime}\right)=\int_{-\infty}^{\infty}\!\!\!\!\!\! dt_{1}dt_{2}\frac{J_{t}\left(t_{1}\right)J_{t}\left(t_{2}\right)}{4\pi^{2}av_{F}}\Theta(t-t_{1})\nonumber \\
 & \times\int_{-\infty}^{\infty}\!\!\!\!\! d\omega\sum_{n}J_{n}^{2}\left(\frac{V_{ac}}{\Omega}\right)e^{-i\omega(t_{1}-t_{2})}G_{aa}^{A}(t_{2},t^{\prime})\nonumber \\
 & \times\left[f\left(\omega-n\Omega+V_{dc}\right)+f\left(\omega+n\Omega-V_{dc}\right)-1\right]\nonumber \\
 & +\int_{-\infty}^{\infty}\!\!\!\!\!\! dt_{1}dt_{2}\frac{J_{a}\left(t_{1}\right)J_{a}\left(t_{2}\right)}{4\pi^{2}av_{F}}\Theta(t-t_{1})\nonumber \\
 & \times\int_{-\infty}^{\infty}\!\!\!\!\! d\omega\left[2f\left(\omega\right)-1\right]e^{-i\omega(t_{1}-t_{2})}G_{aa}^{A}(t_{2},t^{\prime})\:.
\end{align}

The above equations induce the following differential equations 
\begin{align}
 & \!\!\!\left[\partial_{t}+\Gamma_{s}\left(t\right)\right]G_{ba}^{K}(t,t^{\prime})=HG_{aa}^{K}(t,t^{\prime})+H\Lambda_{s}\left(t,t^{\prime}\right)\,,\nonumber \\
 & \!\!\!\left[\partial_{t}+\Gamma_{at}\left(t\right)\right]G_{aa}^{K}(t,t^{\prime})+HG_{ba}^{K}(t,t^{\prime})=\Lambda_{at}\left(t,t^{\prime}\right)\,,\label{eq:solve3}
\end{align}
with $\Lambda_{at/s}\left(t,t^{\prime}\right)=\partial_{t}\Upsilon_{at/s}\left(t,t^{\prime}\right)$.
We define non-interacting Green's functions $g_{at/s}^{K}\left(t,t^{\prime}\right)$
as 
\begin{align}
\left[\partial_{t}+\Gamma_{at/s}\left(t\right)\right]g_{at/s}^{K}(t,t^{\prime}) & =\delta(t-t^{\prime})\,,\label{eq:green0-1-1}
\end{align}
so 
\begin{align}
 & g_{at}^{K}\left(t,t^{\prime}\right)=\Theta\left(t-t^{\prime}\right)\exp\left[-\frac{\Gamma_{at}}{2}\left(t-t^{\prime}\right)\right]\nonumber \\
 & \times\exp\left[-\frac{\Gamma_{at}}{4\Omega_{1}}\left[\sin\left(2\Omega_{1}t\right)-\sin\left(2\Omega_{1}t^{\prime}\right)\right]\right]\:,\nonumber \\
 & g_{s}^{K}\left(t,t^{\prime}\right)=\Theta\left(t-t^{\prime}\right)\exp\left[-\left(\Gamma_{s0}+\frac{\Gamma_{s}}{2}\right)\left(t-t^{\prime}\right)\right]\nonumber \\
 & \times\exp\left[-\frac{2\sqrt{\Gamma_{s0}\Gamma_{s}}}{\Omega_{1}}\left[\sin\left(\Omega_{1}t\right)-\sin\left(\Omega_{1}t^{\prime}\right)\right]\right]\nonumber \\
 & \times\exp\left[-\frac{\Gamma_{s}}{4\Omega_{1}}\left[\sin\left(2\Omega_{1}t\right)-\sin\left(2\Omega_{1}t^{\prime}\right)\right]\right]\:.\label{eq:gK0}
\end{align}
We then have 
\begin{align}
 & \!\!\!\!\! G_{ba}^{K}\!\left(t,t^{\prime}\!\right)\!=\! H\!\int_{-\infty}^{\infty}\!\!\!\!\!\!\! dt_{1}\! g_{s}^{K}\!\left(t,t_{1}\!\right)\!\left[\! G_{aa}^{K}\!\left(t_{1},t^{\prime}\!\right)\!+\!\Lambda_{s}\!\left(t_{1},t^{\prime}\!\right)\!\right]\,,\nonumber \\
 & \!\!\!\!\!\!\int_{-\infty}^{\infty}\!\!\!\!\!\!\! dt_{1}\!\left[\!\left[g_{at}^{K}\!\left(t,t_{1}\!\right)\!\right]^{-1}\!\!\!\!+\! H^{2}\! g_{s}^{K}\!\left(t,t_{1}\!\right)\!\right]\! G_{aa}^{K}\left(t_{1},t^{\prime}\right)\!=\!\Lambda_{at}\!\left(t,t^{\prime}\right)\:.\label{eq:solve2-1}
\end{align}

We realize that the non-interacting Green's functions $g_{at/s}^{A/K}\left(t,t^{\prime}\right)$
oscillate on time with the frequency $\Omega_{1}=\Omega/p$. We cannot
apply any cut-off for the boundary of the integrals in the equations
(\ref{eq:solve2}) and (\ref{eq:solve2-1}). It means we cannot solve
these equations exactly. However, $g_{at/s}^{A/K}\left(t,t^{\prime}\right)$
can be considered oscillating around smooth averaging functions $\overline{g_{at/s}^{A/K}\left(\tau,T\right)}$with
$T\gg\Omega_{1}^{-1}$. It is discussed in detail in the following
subsection.

\subsection{The Averages:}

In order to find the Majorana Green's functions $G_{aa}^{A}\left(t,t^{\prime}\right)$
and $G_{ba}^{K}\left(t,t^{\prime}\right)$, one needs to solve equations
(\ref{eq:solve2}) and (\ref{eq:solve2-1}). In this subsection, we
represent our averaging approximation method.

\subsubsection{Advanced Green's function $\overline{G_{aa}^{A}(\omega)}$:}

The $g_{at/s}^{A}\left(t,t^{\prime}\right)$ in equations (\ref{eq:gA0})
can be expressed as functions of the Bessel function products, for
instance, 
\begin{align}
g_{at}^{A}\left(t,t^{\prime}\right) & =-\Theta(t^{\prime}-t)\sum_{m,n=-\infty}^{\infty}\!\!\!\!\!\!\! J_{n}\left(-i\frac{\Gamma_{at}}{4\Omega_{1}}\right)J_{m}\left(-i\frac{\Gamma_{at}}{4\Omega_{1}}\right)\nonumber \\
 & \times e^{\frac{\Gamma_{at}}{2}\left(t-t^{\prime}\right)}e^{i2\Omega_{1}\left(nt-mt^{\prime}\right)}.
\end{align}
We change time variables such as $\tau=t-t^{\prime}\,,\; T=\left(t+t^{\prime}\right)/2$
. Because the Green's functions $g_{at/s}^{A}(\tau,T)$ oscillate
on time with period $2\pi\Omega_{1}^{-1}$, we average them over time
$T$ in a period of $2\pi/\Omega_{1}$. We then write it in Fourier
space as

\begin{equation}
\overline{g_{at}^{A}(\omega)}=\sum_{n=-\infty}^{\infty}\!\!\!\!\!\!\! J_{n}^{2}\left(-i\frac{\Gamma_{at}}{4\Omega_{1}}\right)\frac{i}{\omega+2n\Omega_{1}-i\frac{\Gamma_{at}}{2}}\,.
\end{equation}
Similarly, 
\begin{align}
\overline{g_{s}^{A}(\omega)} & =\!\!\!\sum_{m,n,k=-\infty}^{\infty}\!\!\!\!\!\!\! J_{m-2n+2k}\left(-i\frac{2\sqrt{\Gamma_{s0}\Gamma_{s}}}{\Omega_{1}}\right)\nonumber \\
 & \times J_{m}\left(-i\frac{2\sqrt{\Gamma_{s0}\Gamma_{s}}}{\Omega_{1}}\right)J_{n}\left(-i\frac{\Gamma_{s}}{4\Omega_{1}}\right)J_{k}\left(-i\frac{\Gamma_{s}}{4\Omega_{1}}\right)\nonumber \\
 & \times\frac{i}{\omega+\left(m+2k\right)\Omega_{1}-i\left(\Gamma_{s0}+\frac{\Gamma_{s}}{2}\right)}\,.
\end{align}
We thus have 
\begin{equation}
\overline{G_{aa}^{A}(\omega)}=\frac{-i}{\overline{g_{at}^{A}(\omega)}^{-1}+H^{2}\overline{g_{s}^{A}(\omega)}}\,.\label{eq:Gaa}
\end{equation}

\subsubsection{Keldysh Green's function $\overline{G_{ba}^{K}(\omega)}$:}

In the same technique, we obtain the average non-interacting Green's
functions 
\begin{equation}
\overline{g_{at}^{K}(\omega)}=\sum_{n=-\infty}^{\infty}\!\!\! J_{n}^{2}\left(i\frac{\Gamma_{at}}{4\Omega_{1}}\right)\frac{i}{\omega+2n\Omega_{1}+i\frac{\Gamma_{at}}{2}}\,,
\end{equation}
 
\begin{align}
\overline{g_{s}^{K}(\omega)} & =\!\!\!\sum_{m,n,k=-\infty}^{\infty}\!\!\!\!\!\!\! J_{m-2n+2k}\left(i\frac{2\sqrt{\Gamma_{s0}\Gamma_{s}}}{\Omega_{1}}\right)\nonumber \\
 & \times J_{m}\left(i\frac{2\sqrt{\Gamma_{s0}\Gamma_{s}}}{\Omega_{1}}\right)J_{n}\left(i\frac{\Gamma_{s}}{4\Omega_{1}}\right)J_{k}\left(i\frac{\Gamma_{s}}{4\Omega_{1}}\right)\nonumber \\
 & \times\frac{i}{\omega+\left(m+2k\right)\Omega_{1}+i\left(\Gamma_{s0}+\frac{\Gamma_{s}}{2}\right)}\:.
\end{align}

We also average $\Lambda_{at/s}\left(t,t^{\prime}\right)$ and obtain
$\overline{\Lambda_{at/s}\left(\omega\right)}$ as 
\begin{align}
 & \!\!\!\!\!\!\!\overline{\Lambda_{at}\left(\omega\right)}=G_{aa}^{A}\left(\omega\right)\!\left\{ \Gamma_{a}\!\left[f\left(\omega-\Omega/p\!\right)\!+f\left(\omega+\Omega/p\!\right)\!-1\!\right]\right.\nonumber \\
 & \!\!\!\!\!\!\!\left.+\!\frac{\Gamma_{t}}{2}\!\sum_{n}\!\! J_{n}^{2}\!\left(\!\frac{V_{ac}}{\Omega}\!\right)\left[f\left(\omega-\left(n+1/p\right)\Omega+V_{dc}\!\right)\right.\right.\nonumber \\
 & \!\!\!\!\!\!\!\left.\left.+f\!\left(\omega+\!\left(n-\!1/p\right)\Omega-V_{dc}\!\right)\!+f\!\left(\!\omega-\!\left(n-\!1/p\right)\Omega+V_{dc}\!\right)\right.\right.\nonumber \\
 & \!\!\!\!\!\!\!\left.\left.+f\left(\omega+\left(n+1/p\right)\Omega-V_{dc}\right)-2\right]\right\} \:,
\end{align}
\begin{align}
 & \overline{\Lambda_{s}\left(\omega\right)}=g_{s}^{A}\left(\omega\right)G_{aa}^{A}\left(\omega\right)\left\{ 2\Gamma_{s0}\left[2f\left(\omega\right)-1\right]\right.\nonumber \\
 & \left.+\Gamma_{s}\left[f\left(\omega-\Omega/p\right)+f\left(\omega+\Omega/p\right)-1\right]\right\} \:.
\end{align}
In this limit, $G_{ba}^{K}(t,t^{\prime})$ only depends on $\tau=t-t^{\prime}$and
its Fourier transform is

\begin{equation}
\overline{G_{ba}^{K}(\omega)}=\frac{H\overline{g_{s}^{K}\left(\omega\right)}\left[\overline{g_{at}^{K}\left(\omega\right)}\overline{\Lambda_{at}\left(\omega\right)}+\overline{\Lambda_{s}\left(\omega\right)}\right]}{1+H^{2}\overline{g_{at}^{K}\left(\omega\right)}\overline{g_{s}^{K}\left(\omega\right)}}\:.\label{eq:GKba}
\end{equation}

The meaning of our averaging method is that we have smoothed the fine
oscillations of the Green's functions. It works well in the high frequency
regime, namely, $T\gg\Omega_{1}^{-1}$. It offers the opportunity
to compute the physical observables non-perturbatively. Our results
should be close to the exact solution while the results from the perturbation
calculation cannot be. Any cut-off in the perturbation calculation
will induce a loosing of some contributions in the physical observables.
However, our averaging method cannot be applied for the adiabatic
limit. The adiabatic regime should be considered separately, maybe
as in Ref. \cite{Adiabatic_pumping}. In the next section, we calculate
the differential conductance and magnetic impurity susceptibility
by using this averaging method.

\section{average physical observables \label{sec:average-physical-observables}}

In this section we will compute average differential conductance and
magnetic susceptibility, and discuss the results. We notice readers
that the word ``average'' in this paper mostly refers our averaging
method.

\subsection{Average charge current and average differential conductance}

We are first interested in calculating the time averaged charge current
through the junction $\langle I_{c}\rangle$ because the time averaged
differential conductance $G=d\langle I_{c}\rangle/dV_{dc}$ is accessible
experimentally \cite{Experiment}. We compute it by using the Keldysh
non-equilibrium Green's function technique. The current at a time
$t$ can be written in the form 
\begin{equation}
I_{c}\left(t\right)=-\frac{eJ_{t}(t)}{4\sqrt{\pi a}}\Re\sum_{k}G_{fka}^{K}\left(t,t\right)\:,
\end{equation}
where we have defined the Green's function $G_{fka}^{K}\left(t,t\right)$
as $G_{fka}\left(t,t^{\prime}\right)=-i\left\langle T_{K}\left\{ \left(c_{fk}^{\dagger}\left(t\right)+c_{fk}\left(t\right)\right)\hat{a}\left(t^{\prime}\right)\right\} \right\rangle $.
The current is then expressed as $I_{c}\left(t\right)=-\left(eJ_{t}(t)/8\pi a\right)\Re\int_{-\infty}^{\infty}dt_{1}J_{t}(t_{1})m_{f}^{K}\left(t,t_{1}\right)G_{aa}^{A}\left(t_{1},t\right)$.
As we apply the average condition to the Green's functions, the current
at a time $t$ means ``average'': $\overline{I_{c}\left(t\right)}\sim\overline{m_{f}^{K}\left(t,t_{1}\right)G_{aa}^{A}\left(t_{1},t\right)}\sim\overline{m_{f}^{K}\left(t,t_{1}\right)}\overline{G_{aa}^{A}\left(t_{1},t\right)}$
then 
\begin{align}
\overline{I_{c}\left(t\right)} & =\frac{J_{t}\left(t\right)}{8\pi^{2}av_{F}}\Im\int_{-\infty}^{\infty}\!\!\!\!\!\!\! dt_{1}J_{t}\left(t_{1}\right)\int_{-\infty}^{\infty}\!\!\!\!\! d\omega\sum_{n}J_{n}^{2}\left(\frac{V_{ac}}{\Omega}\right)\nonumber \\
 & \times\left[f\left(\omega+n\Omega-V_{dc}\right)-f\left(\omega-n\Omega+V_{dc}\right)\right]\nonumber \\
 & \times e^{-i\omega\left(t-t_{1}\right)}G_{aa}^{A}\left(t_{1},t\right)\;.
\end{align}
One should notice that we calculate the time averaged differential
conductance, we therefore can apply our average Green's function method,
which has been discussed in section \ref{sec:Average-Green's-functions}.
The Majorana Green's functions $G_{aa}^{A}\left(t_{1},t\right)$ is
calculated averagely in the above section with its Fourier image is
shown in equation (\ref{eq:Gaa}). As discussed above, the current
should oscillate on time with period $2\pi\Omega^{-1}$. It is possible
to take the average of the current over time $t$ in a period of $2\pi p\Omega^{-1}$.
So the average current is 
\begin{align}
 & \overline{I_{c}}=\frac{\Omega}{2\pi p}\int_{0}^{2\pi p/\Omega}\!\!\!\!\! dt\,\overline{I_{c}\left(t\right)}\nonumber \\
 & =\frac{\Gamma_{t}}{8\pi}\Im\int_{-\infty}^{\infty}\!\!\!\!\! d\omega\sum_{n}J_{n}^{2}\left(\frac{V_{ac}}{\Omega}\right)\nonumber \\
 & \times\left\{ \left[f\left(\omega+\left(n-1/p\right)\Omega-V_{dc}\right)\right.\right.\nonumber \\
 & \left.\left.-f\left(\omega-\left(n+1/p\right)\Omega+V_{dc}\right)\right.\right.\nonumber \\
 & \left.\left.+f\left(\omega+\left(n+1/p\right)\Omega-V_{dc}\right)\right.\right.\nonumber \\
 & \left.\left.-f\left(\omega-\left(n-1/p\right)\Omega+V_{dc}\right)\right]G_{aa}^{A}\left(\omega\right)\right\} \:.\label{eq:current}
\end{align}
By taking the derivative of the time averaged current in equation
(\ref{eq:current}) with respect to the dc voltage, we obtain the
time averaged differential conductance. The conductance and the magnetic
susceptibility behaviors remain when the temperature is varied from
absolute zero to a very small temperature. If the temperature is kept
increasing much below the Kondo temperature, the conductance and the
susceptibility behaviors are broadened and then smeared. For convenient
computation, we consider the time averaged differential conductance
at the absolute zero 
\begin{align}
 & \overline{G}=\frac{\Gamma_{t}}{8\pi}\Im\sum_{n=-\infty}^{\infty}J_{n}^{2}\left(\frac{V_{ac}}{\Omega}\right)\nonumber \\
 & \times\left\{ \overline{G_{aa}^{A}\left(V_{dc}-\left(n-1/p\right)\Omega\right)}\right.\nonumber \\
 & \left.+\overline{G_{aa}^{A}\left(-V_{dc}+\left(n+1/p\right)\Omega\right)}\right.\nonumber \\
 & \left.\overline{G_{aa}^{A}\left(V_{dc}-\left(n+1/p\right)\Omega\right)}\right.\nonumber \\
 & \left.+\overline{G_{aa}^{A}\left(-V_{dc}+\left(n-1/p\right)\Omega\right)}\right\} \;,\label{eq:G_result}
\end{align}
where the differential conductance is defined as $G=d\left[\frac{\Omega}{2\pi p}\int_{0}^{2\pi p/\Omega}I_{c}\left(t\right)dt\right]/dV_{dc}$.
We will discuss the behavior of the average differential conductance
as a function of magnetic field and source-drain voltage.

\subsection{Average impurity magnetization and susceptibility}

While the experimentalists are interested in measuring average differential
conductance, the theorists are interested in calculating magnetic
susceptibility as well because both observables provide direct information
about the magnetic impurity of a SET. The local magnetic susceptibility
-- the response to a magnetic field coupled to the impurity spin only
-- is the derivative of the magnetization, which is calculated from
the following formula

\begin{equation}
M\left(H\right)=\frac{\mu_{B}g_{i}}{4\pi}\!\int_{-\infty}^{\infty}\!\!\!\!\! d\omega G_{ba}^{K}\!\left(\omega\right)\:.
\end{equation}
The Majorana Green's function $G_{ba}^{K}\left(\omega\right)$ is
calculated averagely in the above section with its Fourier image shown
in equation (\ref{eq:GKba}). The magnetization is understood as the
``average'' one and is expressed as 
\begin{equation}
M\left(H\right)=\frac{\left(\mu_{B}g_{i}\right)H}{4\pi}\!\!\int_{-\infty}^{\infty}\!\!\!\!\!\! d\omega\!\frac{\overline{g_{s}^{K}\left(\omega\right)}\!\left[\!\overline{g_{at}^{K}\left(\omega\right)}\overline{\Lambda_{at}\left(\omega\right)}\!+\!\overline{\Lambda_{s}\left(\omega\right)}\!\right]}{1+H^{2}\overline{g_{at}^{K}\left(\omega\right)}\overline{g_{s}^{K}\left(\omega\right)}}\:.\label{eq:magnetization}
\end{equation}
From formula (\ref{eq:magnetization}), when the system is applied
either a constant voltage bias $V_{sd}=V_{dc}$ or a time dependent
one $V_{sd}=V_{dc}+V_{ac}\cos\left(\Omega t\right)$ and the Kondo
couplings are time independent, again, we re-obtain the results in
the SH theory \cite{SchillerHershfield-PRB,SchillerHershfield}.

We compute the impurity susceptibility as $\chi\left(H\right)=\mu_{B}g_{i}\partial_{H}M\left(H\right)$
and discuss its behavior in the following subsection.

\subsection{Results and discussions}

What we show below is just a sample of some plots to give a flavor
of our results. In all plots, the horizontal axis is the magnetic
energy $H=\mu_{B}g_{i}B$, $B$ is the amplitude of applied magnetic
field, the vertical axis is conductance $G/(e^{2}/\hbar)$ or magnetic
susceptibility $\chi/(g_{i}\mu_{B})^{2}$. We choose $\Gamma_{s0}=1$
as the energy scale and it can be considered Kondo temperature. It
means the other parameter values are chosen in comparing with $\Gamma_{s0}$.
The results obtained at very small temperature and at absolute zero
are similar. For efficient numerical calculation we choose to work
with conductance at absolute zero and susceptibility at very small
temperature ($T=0.05$).

\begin{figure}[t]
\includegraphics[width=7cm]{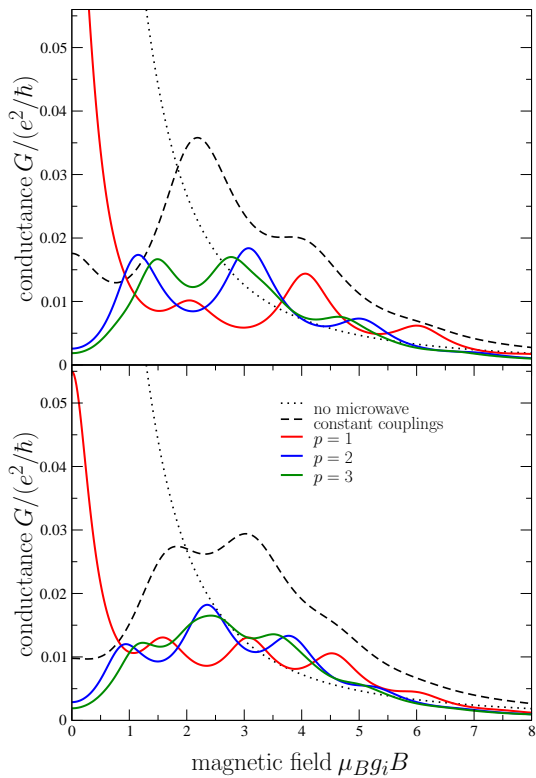} \caption{(Color online) $G\left(H\right)$ are plotted for $V_{ac}=4,\:\Gamma_{at}=0.3,\:\Gamma_{s}=0.05,\:\Gamma_{t}=0.25,\: V_{dc}=0$,
and $\hbar\Omega=2$ in the upper panel and $\hbar\Omega=1.5$ in
the lower one. The dotted black curve is without microwaves; Kondo
peak at $\mu_{B}g_{i}B=0$. The dashed black curve is with ac voltage,
constant couplings: $J_{\perp}^{\alpha\beta}=J_{\perp0}^{\alpha\beta}+J_{\perp1}^{\alpha\beta}$;
Kondo satellites at $\mu_{B}g_{i}B=n\hbar\Omega$; the highest peak's
position depends on $eV_{ac}/\hbar\Omega$. When both voltage and
couplings are oscillating, peaks occur at $\mu_{B}g_{i}B=\left(n\pm1/p\right)\hbar\Omega$;
the solid red, blue, green curves are corresponding for the $p=1,\:2,\:3$.
The peak splitting due to time-dependent couplings can be seen clearly
at the highest peak.}

\label{f.1} 
\end{figure}

\begin{figure}[t]
\includegraphics[width=7cm]{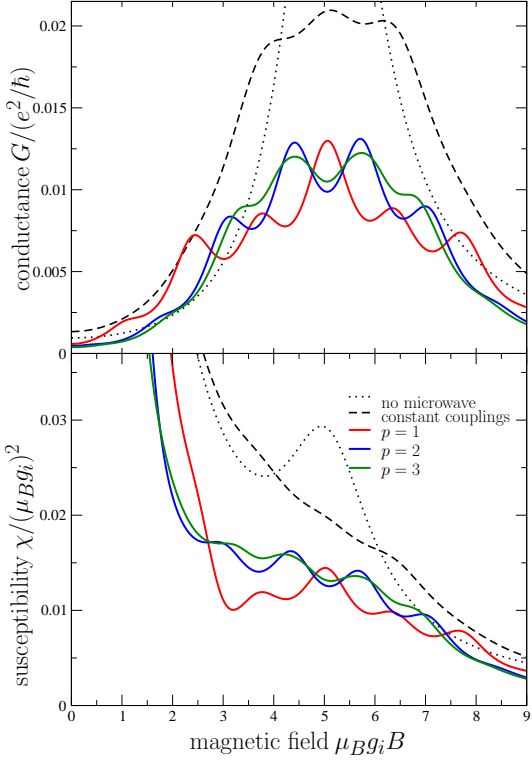} \caption{(Color online) $G\left(H\right)$ and $\chi\left(H\right)$ are plotted
in the upper and lower panels correspondingly for $\hbar\Omega=1.35,\: eV_{ac}=2,\: V_{dc}=5,$
$\Gamma_{at}=0.3,\:\Gamma_{s}=0.1,\:\Gamma_{t}=0.25$. The dotted
black curve is without microwaves; Kondo peak at $\mu_{B}g_{i}B=eV_{dc}$.
The dashed black curve is with ac voltage, constant couplings: $J_{\perp}^{\alpha\beta}=J_{\perp0}^{\alpha\beta}+J_{\perp1}^{\alpha\beta}$;
Kondo satellites at $\mu_{B}g_{i}B=eV_{dc}\pm n\hbar\Omega$. When
both voltage and couplings are oscillating, peaks occur at $\mu_{B}g_{i}B=eV_{dc}\pm\left(n\pm1/p\right)\hbar\Omega$;
the solid red, blue, green curves are corresponding for the $p=1,\:2,\:3$. }

\label{f.3} 
\end{figure}

In figures \ref{f.1} and \ref{f.3}, conductance $G$ and susceptibility
$\chi$ are plotted as functions of magnetic field $H$ for different
situations: no microwave (dotted black curves), time dependent voltage
(dashed black curves), and both time dependent voltage and couplings
with different values of $p$ (red curves for $p=1$, blue curves
for $p=2$, and green curves for $p=3$). When we compare the conductance
with and without the effect of microwaves, we find that microwave
irradiation reduces strongly the conductance. Because we consider
a SET in a special situation of couplings, we cannot compare our results
with the case where we only consider oscillation in voltage $V_{ac}\left(t\right)$,
no time oscillation applied on couplings as we discussed in section
\ref{sec:Kondo-model} (no current in this situation in equilibrium).
However, we can compare our result with the case in which we only
consider oscillation in voltage $V_{ac}\left(t\right)$ and constant
couplings such as $J_{\perp}^{\alpha\beta}=J_{\perp0}^{\alpha\beta}+J_{\perp1}^{\alpha\beta}$
so $J_{t},\: J_{s0}+J_{s},\: J_{a}$. The side-band effect of microwaves
with the peaks appearing at $\mu_{B}g_{i}B=eV_{dc}\pm n\hbar\Omega$
($n\in\mathbb{Z}$) due to the oscillation of source-drain voltage
$V_{ac}\left(t\right)$ \cite{Tien_Gordon,Hettler_Schoeller,SchillerHershfield,Avishai,Lopez,Nordlander,Kaminski,sideband_exp}.
As it is mentioned in Ref. \cite{sideband_exp} the photon-induced
satellites appear quite delicate, we deeply investigate the effect
of ratio $eV_{ac}/\hbar\Omega$. The side-band peaks only appear when
$eV_{ac}/\hbar\Omega$ is big enough compare to $1$ (but $eV_{ac}\sim\hbar\Omega$,
as well as $\hbar\Omega\sim k_{B}T_{K}$ and $\hbar\Omega>k_{B}T_{K}$
in order to avoid decoherence), and this ratio determines the height
of side-band peaks. For instance, in the upper panel of figure \ref{f.1}
plotted for $1<eV_{ac}/\hbar\Omega\leq2$, we see the highest peak
at $\mu_{B}g_{i}B=\hbar\Omega$ besides the trivial main peak at $\mu_{B}g_{i}B=0$.
The strength of microwaves is small compared to the strength of the
magnetic field, which reduces strongly Kondo conductance. When we
increase the oscillation amplitude, microwave effect becomes stronger
than the effect of the magnetic field, it reduces strongly the peak
at $\mu_{B}g_{i}B=0$ and increases the satellite peak at $\mu_{B}g_{i}B=2\hbar\Omega$
as we see in the lower panel for $eV_{ac}/\hbar\Omega>2$. One can
say the highest satellite peak position depends on the ratio $eV_{ac}/\hbar\Omega$.
However, it is impossible to see the highest satellite peak at $\mu_{B}g_{i}B=3\hbar\Omega$
due to the strong suppression by both magnetic field and voltage.
This result suggests to experimentalists an efficient method to calibrate
the oscillation voltage amplitude $V_{ac}$.

We now discuss the second effect of microwaves. In fact, the oscillation
of Kondo couplings comes from the oscillation of source-drain and
gate voltages. It can be considered as an indirect effect. In figure
\ref{f.1} the dashed black curve is plotted for the constant couplings,
the red, blue, and green curves are plotted for the time dependent
couplings with $p=1,\:2,\:3$ correspondingly. We find that when the
coupling parameters oscillate with frequency $\Omega_{1}=\Omega/p$
with $p\in\mathbb{N}$, each main peak splits into two peaks. Since
the energy is conserved, when a peak is split into two peaks, the
height of these two peaks relate to the height of original peak. The
distance between an original peak and its split peak is $d_{peak}=\hbar\Omega/p$.
This is understood mathematically by investigating the Green's function
$G_{aa}^{A}(\omega)$. The difference between the coupling frequency
and the applied frequency is described by an integer number $p\geq1$.
However, $p$ cannot be too big because the slow oscillation in couplings
can be considered as constant compared to the fast oscillation of
voltage. When $p=1$, the couplings oscillate on time with the same
frequency as microwaves, the satellite peaks split into two peaks
with the distance $d_{peak}=\hbar\Omega$ equal to the satellite peak
distance. So for $p=1$, the peaks remain but their height changes
drastically due to the re-contribution of energy through the peak
splitting. This peak splitting effect can be seen clearly if $p=2$,
or $p=3$. As we have explained the number $p$ can also be bigger
than $1$ due to the delay of phase between the source-drain voltage
and gate voltage. The peak splitting can be seen clearly at the highest
satellite peak. When we increase $p$ the splitting distance $d_{peak}$
decreases, the splitting becomes weak then approaches the limit in
which the couplings are constant. This result explains well a possibility
in experimental result in which one sees peaks appearing at a distance
$\mu_{B}g_{i}B<\hbar\Omega$ in the regime of weak magnetic field.
We predict $p$ is a frequency dependent systematic parameter. It
should be checked carefully in experimentation. For the case $p=1$,
we can check the height of the satellite peaks to confirm the two
fold effect of microwave on voltage and couplings.

\begin{figure}[t]
\includegraphics[width=7cm]{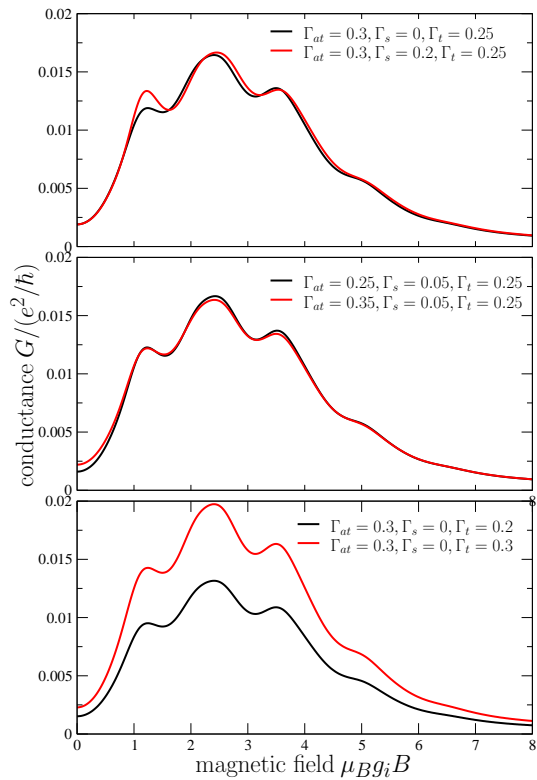} \caption{(Color online) $G\left(H\right)$ are plotted for $V_{ac}=4,\:\hbar\Omega=1.5,\:,p=3,\: V_{dc}=0$,
and different set of energy scales $\Gamma_{i}$. In the top panel,
$\Gamma_{at}=0.3,\:\Gamma_{t}=0.25$ is fixed, $\Gamma_{s}$ is varied:
$\Gamma_{s}=0$ for the black curve, $\Gamma_{s}=0.2$ for the red
curve. In the middle panel, $\Gamma_{s}=0.05,\:\Gamma_{t}=0.25$ is
fixed, $\Gamma_{at}$ is varied: $\Gamma_{at}=0.25$ for the black
curve, $\Gamma_{s}=0.35$ for the red curve. In the bottom panel,
$\Gamma_{at}=0.3,\:\Gamma_{s}=0$ is fixed, $\Gamma_{t}$ is varied:
$\Gamma_{t}=0.2$ for the black curve, $\Gamma_{t}=0.3$ for the red
curve. Notice conductance amplitude strongly depends on $\Gamma_{t}$
.}

\label{f.2} 
\end{figure}

Besides, from figure \ref{f.2}, we find the main characteristic of
$G\left(H\right)$ lightly depends on scales $\Gamma_{i}$, ($i=at,s,t$).
$\Gamma_{i}$ are chosen much smaller than $\Gamma_{s0}$ because
the microwave effects on the couplings should be small compared to
the existing couplings. We vary each energy scale $\Gamma_{s},\:\Gamma_{at},\:\Gamma_{t}$
in the top, middle, and bottom panel correspondingly while other parameters
are fixed. $\Gamma_{t}\sim\left(J_{\perp}^{LR}\right)^{2}$concerns
the coupling between two leads and it controls the spin-flip tunneling
of electrons through QD. $\Gamma_{s}\sim\left[J_{\perp}^{LL}+J_{\perp}^{RR}\right]^{2}$concerns
the sum of separated couplings in each lead and is small compared
to $\Gamma_{s0}$. $\left(J_{\perp}^{LL}-J_{\perp}^{RR}\right)^{2}$
contributes to $\Gamma_{at}$. Both energy scales $\Gamma_{at}$ and
$\Gamma_{s}$ affect slightly the $G\left(H\right)$ behavior. The
red curve in the bottom panel is plotted for the case in which modulation
of microwave only affects the transverse couplings, so $\Gamma_{at}=0.3,\:\Gamma_{s}=0,\:\Gamma_{t}=0.3$.
The amplitude of conductance strongly decreases when we decrease $\Gamma_{t}$.
It vanishes when $\Gamma_{t}=0$. Thus, any small effect of microwave
irradiation drives the system away from the uncoupled isotropic two-channel
Kondo situation.

In figure \ref{f.3}, the magnetic susceptibility is plotted as a
function of magnetic field at temperature $T=0.05$. As previously
stated from $G\left(H\right)$ characteristic, dc voltage $V_{dc}$
also splits the satellite peaks \cite{Hettler_Schoeller} when it
is smaller than the ac voltage $V_{ac}$. It may cause confusion to
recognize the peak splitting due to the oscillations of couplings
in experiment if experimentalists cannot avoid a small contribution
of $V_{dc}$. The fact that $V_{dc}$ splits peaks by a distance $\pm eV_{dc}$
is understood in non-equilibrium Kondo physics of a QD without microwave
irradiation \cite{SchillerHershfield-PRB}. Significantly, if $V_{dc}$
is bigger than $V_{ac}$, we see the peak splitting due to the oscillations
of couplings. If $V_{dc}=0$, we see the satellite peaks dominate
at $\mu_{B}g_{i}B=0$. The other peaks at $\mu_{B}g_{i}B=k\hbar\Omega$
($k=\pm1,\:\pm2,\:\ldots$) are very weak. To study the satellite
peak splitting due to coupling oscillation, we choose $V_{dc}>V_{ac}$.
We find all susceptibility-magnetic field and conductance-magnetic
field characteristics are the same. It is evident that the relation
between susceptibility and conductance shown in formula (8.9) of Ref.
\cite{SchillerHershfield-PRB} can be generalized when the system
is irradiated by microwaves.

\begin{figure}[t]
\includegraphics[width=7cm]{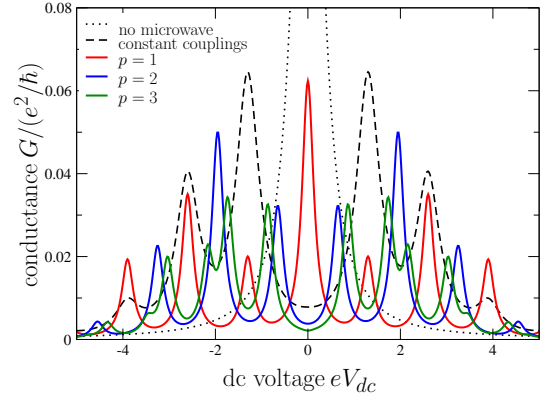} \caption{(Color online) $G\left(V_{dc}\right)$is plotted for $\hbar\Omega=1.3,\: eV_{ac}=3,\:\Gamma_{at}=0.3,\:\Gamma_{s}=0.05,\:\Gamma_{t}=0.2,\: H=0$.
The dotted black curve is without microwaves; Kondo peak at $eV_{dc}=0$.
The dashed black curve is with ac voltage, constant couplings; Kondo
satellites at $eV_{dc}=\pm n\hbar\Omega$. When both voltage and couplings
are oscillating, peaks occur at $eV_{dc}=\pm\left(n\pm1/p\right)\hbar\Omega$;
the solid red, blue, green curves are corresponding for the $p=1,\:2,\:3$. }

\label{f.4} 
\end{figure}

\begin{figure}[t]
\includegraphics[width=7cm]{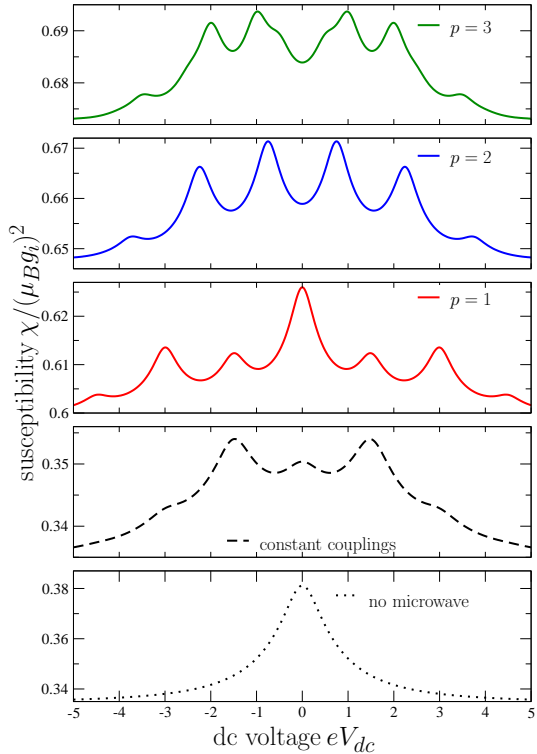} \caption{(Color online) $\chi\left(V_{dc}\right)$ is plotted for $\hbar\Omega=1.5,\: eV_{ac}=2.5,\:\Gamma_{at}=0.3,\:\Gamma_{s}=0.1,\:\Gamma_{t}=0.25,\: H=0$.
The dotted black curve is without microwaves; Kondo peak at $eV_{dc}=0$.
The dashed black curve is with ac voltage, constant couplings; Kondo
satellites at $eV_{dc}=\pm n\hbar\Omega$. When both voltage and couplings
are oscillating, peaks occur at $eV_{dc}=\pm\left(n\pm1/p\right)\hbar\Omega$;
the solid red, blue, green curves are corresponding for the $p=1,\:2,\:3$. }

\label{f.5} 
\end{figure}

The differential conductance $G$ and magnetic susceptibility $\chi$
are also plotted as functions of dc voltage $V_{dc}$ in figures \ref{f.4}
and \ref{f.5}. In order to investigate microwave irradiation effect,
we consider the case in which no magnetic field is applied. Again,
we find the two-fold peak splitting (Kondo satellite splitting) due
to both voltage and Kondo coupling oscillations. There is more challenge
for experimentalists if we are in the cases $p=1$ and $p=2$. The
distance between peaks is still $\hbar\Omega$. For $p=1$, the positions
of peaks are the same as those in the case when only voltage oscillates,
but the peak heights are re-contributed. For $p=2$, both the positions
and heights of peaks are re-contributed, satellite peaks are shifted
by $\hbar\Omega/2$. For $p\geq3$, our interesting results show that
one has a chance to see two peaks, whose distance is much smaller
than $\hbar\Omega$. For instance, in the green curves in figures
\ref{f.4} and \ref{f.5}, which are plotted for $p=3$, one can find
two peaks which are split from a satellite peak at a distance of $2\hbar\Omega/3$.
Moreover, one can also find two peaks at a distance of $\hbar\Omega/3$.
They are the peaks, that are split from the consecutive satellite
peaks.

\section{Conclusions \label{sec:Conclusions}}

In this paper, we have investigated the effect of microwave irradiation
on a SET in a magnetic field. The frequency of the microwaves is $\Omega$.
The system is described by a non-equilibrium Kondo model in a specific
point in the parameter space, where the Hamiltonian $\mathcal{H}$
can be diagonalized. Besides the oscillation in the source-drain voltage,
the oscillation in the Kondo couplings is investigated for a specific
situation. The Kondo couplings $J_{\perp}^{\alpha\beta}(t)$ relate
to the source-drain voltage and the gate voltage, which oscillate
with frequency $\Omega$. The couplings thus oscillate on time with
the same frequency or its harmonics. However, one can think about
an appropriate experimental set-up, in which the Kondo couplings $J_{\perp}^{\alpha\beta}(t)$
oscillate with a frequency $\Omega_{1}=\Omega/p$, $p\in\mathbb{N}$.
The time dependent parts in $J_{\perp}^{\alpha\beta}(t)$ drive the
system away from the isotropic two channel Kondo problem to the anisotropic
one though their amplitudes are small compared to the time independent
parts. Due to the oscillations in the input parameters, the interacting
Green's functions oscillate on time with a frequency of $\hbar\Omega/p$;
thus the problem cannot be solved exactly. However, these fast oscillations
allow one to average Green's functions and observables in a period
$2\pi p/\Omega$. We have proposed a non-perturbative approximation,
in which we have smoothed the fine fast oscillations around its average
form. Thus, the Green's functions and the observables are averaged.
Our averaging method gives the results, which are closed to the exact
one. The higher the frequency, the better the approximation result.
It works well for the cases $\hbar\Omega\gtrsim k_{B}T_{K}\sim\Gamma_{s0}\gg\Gamma_{at/s/t}$
in which we study Kondo satellite peaks. 

The center result of our work is the satellite peak splitting. This
feature happens when oscillation parts are added to the Kondo couplings
in a particular situation. The distance between two peaks, which are
split from a satellite, is $2\hbar\Omega/p$, the distance between
two peaks, which are split from two consecutive satellites, is $\left(p-1\right)\hbar\Omega/p$,
while the distance between two satellites is $\hbar\Omega$. All the
new distances in our results depend on the number $p$, which describes
the difference between the frequency of the Kondo couplings and the
frequency of input microwaves. We see the two close peaks clearly
at small magnetic field in $G\left(H\right),\:\chi\left(H\right)$
characteristics and small dc voltage in $G\left(V_{dc}\right),\:\chi\left(V_{dc}\right)$
characteristics.

Magnetic field and applied dc and ac voltages suppress the Kondo correlation
alternatively. When a dc source-drain voltage $V_{dc}$ is applied,
the equilibrium Kondo peak is split into two peaks \cite{KondoQDnoneq1,KondoQDnoneq2}.
The distance from the original peak to one of the new ones is $d_{peak}=eV_{dc}$.
When a magnetic field $B$ is applied, the Kondo peak is also split
into two peaks with $d_{peak}=\mu_{B}g_{i}B$ \cite{KondoQDnoneq1,KondoQD_Bfield}.
When an ac source-drain voltage is applied, the Kondo peak is expanded
into satellites with $d_{peak}=n\hbar\Omega$ \cite{Hettler_Schoeller,SchillerHershfield,Avishai,Lopez,Kaminski,Nordlander}.
When all above external fields are applied to a SET, the equilibrium
Kondo peak is split into satellites. The position of a satellite is
$\pm eV_{dc}\pm\mu_{B}g_{i}B\pm\left(n\pm1/p\right)\hbar\Omega$ when
the Kondo couplings are considered oscillating on time with frequency
$\hbar\Omega/p$.

Although some works, in which the time dependent exchange interactions
like Kondo couplings $J\left(t\right)$ were considered, have been
done by solving flow equations \cite{Kehrein-1}, the exact solution
for time dependent Kondo model still remains. In our work, we proposed
the averaging method, which is good for high frequency regime. We
predict that another method is needed for the adiabatic regime. Besides,
one can discuss more carefully about the effect of microwave irradiation
on the Kondo channels. Moreover, our work suggests qualitatively experimentalists
set up experiments in which more than one frequency is activated \cite{Experiment}.

\section{Acknowledgments}

We gratefully acknowledge Carlos Bolech and Nayana Shah for the project
suggestion and numerous discussions. The discussions with Andrei Kogan
and Bryan Hemingway on experimental realization and their data sharing
are gratefully acknowledged. We are thankful to Tran Minh Tien for
a fruitful discussion about integral equations and to Philippe Dollfus
for their review. This work is supported by University of Cincinnati.


\begin{thebibliography}{10}
\bibitem{discussion} the different experimental set-ups for different
scenarios of microwave effects have been discussed with Andrei Kogan's
group.
\end{thebibliography}
\end{document}